\documentclass[pra,twocolumn,showpacs,floatfix]{revtex4}
\usepackage{graphicx}
\usepackage{color}
\usepackage{amsmath}
\begin{document}

\title{Ground state and rotational properties of two-dimensional self-bound quantum
droplets}

\author{P. Examilioti$^1$ and G. M. Kavoulakis$^2$}
\affiliation{$^1$Department of Physics, University of Crete, Heraklion, GR-71003, Greece
\\
$^2$Hellenic Mediterranean University, P.O. Box 1939, GR-71004, Heraklion, Greece}

\begin{abstract}

We consider a two-dimensional self-bound quantum droplet, which consists of a mixture 
of two Bose-Einstein condensates. We start with the ground state, and then turn to 
the rotational response of this system, in the presence of an external (harmonic) 
potential. We identify various phases, depending on the atom number, the strength of 
the external confinement and the angular momentum. These include center of mass 
excitation, ghost vortices, as well as vortices of single and multiple quantization. 
According to our results, this is an excellent system for the study of superfluid states. 

\end{abstract}


\maketitle

\section{Introduction}

Self-bound, macroscopic, droplets appear often in nature, in various physical 
systems. Probably the most familiar example is that of water droplets, which form due to the 
surface tension, with their surface behaving as though it is covered by an elastic medium. 
Droplets also appear on microscopic length scales. One familiar example is the atomic nucleus 
\cite{Ben}, while another example is that of nano droplets in liquid Helium \cite{lndr}.

A common feature of the above systems is that they are dense and strongly interacting. 
Recently, in the field of cold atomic gases the existence of such droplets was predicted 
theoretically in a two-component Bose-Einstein condensate, see Refs.\,\cite{Petrov, PA}. 
The novel property of these quantum systems is that -- as opposed to the ones mentioned 
above -- they are very dilute and weakly-interacting and as a result they have attracted 
a lot of attention recently.

The main idea is that by suitable tuning of the inter- and intra-component interaction 
strengths, the mean-field term -- which is proportional to the square of the atom density 
$n$ -- is sufficiently weak and attractive and is comparable with the next-order correction
of the energy, i.e., the Lee-Huang-Yang term \cite{LHY}. This term is repulsive and scales 
as $n^{5/2}$ (in three spatial dimensions). As a result, these two terms balance each other,
giving rise to a system which is self-bound, or, in other words, no trapping potential is 
necessary in order for the atoms which constitute the droplet to bind together. 

In cold atomic gases it is possible to realize also quasi-two and quasi-one dimensional systems. 
This becomes possible by some external confining potential, which acts either in one, or in 
two dimensions. Provided that the quantum of energy due to this potential is much larger than 
all the other energy scales in the problem, the system becomes quasi-two, or quasi-one 
dimensional. In this case, despite some differences, quantum droplets are also possible.

Various groups have managed to realize quantum droplets recently in mixtures of Bose-Einstein 
condensed gases \cite{qd7, qd8, qd8a, gd8b, qd8c}. Another system where droplets have been 
observed is that of single-component gases, with strong dipolar interactions \cite{qd1, qd2, 
qd3, qd4, qd5, qd6}. 

Also, several theoretical studies have been performed recently on quantum droplets, and they
have focused on various interesting problems. One basic problem includes the ground-state 
properties of droplets in three, two, and one spatial dimensions. Other questions include 
dynamic properties of the droplets, e.g., their collective excitations, the dynamic formation 
of droplets, the formation of vortices in droplets, etc. Finally, although the droplets are 
self-bound, some studies have considered the effect of an external potential on them. The list 
of references on these problems is long. Some representative studies include Refs.\,\cite{Petrov, 
PA, th0, th1, th2, th3, th4, th5, th6, th7, th8, th9, th10, th11, th12, th13, th14, th15, th16}. 

Quantum droplets are expected to have the collection of properties associated with 
``superfluidity". It is thus natural to examine their rotational response, very much like 
the problem of rotating nuclei, of rotating Helium nano droplets, of the ``traditional" 
rotating, trapped Bose-Einstein condensates of cold atoms, etc. 

This is actually the subject of the present study. More specifically, we examine below the 
rotational response of a quantum droplet, both in the absence, as well as in the presence of 
an external harmonic potential, in two spatial dimensions. Reference \cite{th3} has studied 
the ground state, and the local stability of multiply-quantized vortex states, in a 
two-dimensional droplet. Since in the present study we are interested in the global minimum 
of the energy, even if, e.g., the multiply-quantized vortex states are stable, still the state 
of lowest energy may be a different one. Finally, Ref.\,\cite{th6} has studied the same problem 
as the one we consider here, i.e, the yrast problem, for some fixed value of the atom number 
and of the trap frequency. Its results are fully consistent with the present ones, since the 
atom number that was considered there is large enough and the trapping potential is sufficiently 
strong, so that the droplet carries its angular momentum via vortex excitation (as seen also 
in the results below).

In what follows below we consider a two-dimensional quantum droplet. We start 
with our model in Sec.\,II, while in Sec.\,III we examine the ground-state properties of 
the droplet, evaluating the order parameter numerically, and variationally. The variational 
results turn out to be very accurate and also allow us to get insight into this problem. 
We then turn to the rotational properties of the droplets. We start in Sec.\,IV with 
the case of no external confinement, arguing that the lowest-energy (yrast) state for 
some fixed value of the angular momentum is always the one that involves center of 
mass excitation. 

Introducing a harmonic potential in Sec.\,V, we demonstrate that vortex excitation 
competes with the center of mass motion. Solving the full problem numerically and 
performing also again variational calculations, we identify the yrast states for 
various values of the atom number and of the trap frequency. We then derive the 
corresponding phase diagram that involves these two parameters, for various values 
of the angular momentum. Finally, we summarize our results and present our conclusions
in Sec.\,VI.

\section{Model}

Let consider a (quasi-)two-dimensional droplet, where three-body losses are 
negligible, \cite{PA, th0, th6}. In general there are two order parameters $\Psi_1$ and 
$\Psi_2$, which correspond to the two components. These satisfy the equations
\begin{eqnarray}
 i \hbar \frac {\partial \Psi_j} {\partial t} = - \frac {\hbar^2} {2 M} \nabla^2 \Psi_j
 + V \Psi_j + g (|\Psi_j|^2 - |\Psi_k|^2) \Psi_j 
 \nonumber \\
 + \alpha (|\Psi_j|^2 + |\Psi_k|^2)
\ln \left( \frac {(|\Psi_j|^2 + |\Psi_k|^2)} {n_0} \right) \Psi_j.
\end{eqnarray}
Here, each order parameter is normalized to the number of atoms in the corresponding 
component. Also, the intraspecies interaction is assumed to be repulsive and the 
interspecies interaction attractive. Furthermore, $M$ is the atom mass (assumed to be 
the same for the two components), $V$ is the external potential, $g$ is the term of 
the usual quadratic nonlinear term, while $\alpha$ is the term that refers to the 
Lee-Huang-Yang correction. Finally, the density $n_0$ is the one of the symmetric 
``flat" state \cite{PA, th0, th6}. 

Here, we consider the ``symmetric" case, where $\Psi_1$ and $\Psi_2$ are equal. 
Clearly one may also consider the more general problem, since, as seen below, already 
this model has a very rich structure. In this case, the problem reduces to that of a 
single order parameter $\Psi/{\sqrt 2} = \Psi_1 = \Psi_2$, which satisfies the equation
\begin{eqnarray}
 i \hbar \frac {\partial \Psi} {\partial t} = - \frac {\hbar^2} {2 M} \nabla^2 \Psi
 + V \Psi + \alpha |\Psi|^2 \ln \left( \frac {(|\Psi|^2} {n_0} \right) \Psi.
\label{qee}
\end{eqnarray}
For a harmonic trapping potential of frequency $\omega$, Eq.\,(\ref{qee}) may be written 
in the more convenient, dimensionless, form
\begin{eqnarray}
 i \frac {\partial \Psi} {\partial t} = \left( - \frac {1} {2} \nabla^2 
 + \frac 1 2 \omega^2 \rho^2 + |\Psi|^2 \ln \left( |\Psi|^2 \right) \right) \Psi.
\label{qeee} 
\end{eqnarray}
The corresponding time-independent equation is,
\begin{equation}
 \left( - \frac 1 2 \nabla^2 + \frac 1 2 \omega^2 \rho^2 + |\Psi|^2 \ln |\Psi|^2 \right)
\Psi = \mu \Psi, 
 \label{nlin}  
\end{equation}
where $\mu$ is the chemical potential, while the energy functional is \cite{PA, th0, th6}
\begin{eqnarray}
  E = \int \left( \frac {1} {2} |\nabla \Psi|^2 + \frac 1 2 \omega^2 \rho^2 |\Psi|^2
  + \frac 1 2 |\Psi|^4 \ln \frac {|\Psi|^2} {\sqrt{e}} \right) \, d^2 \rho.
\nonumber \\
\label{funnc}
\end{eqnarray}
The interesting feature of the above functional is the nonlinear term, which may change sign, 
depending on the value of the density.

\section{Ground state}

We have thus solved Eq.\,(\ref{nlin}) using the method of imaginary-time propagation
\cite{imag}. The ground-state energy for various values of $N$ which comes from this
calculation is shown in Fig.\,1, as the solid curve. Within the Thomas-Fermi approximation 
\cite{th3}, which is valid in the limit of large atom numbers $N$, the kinetic energy is 
negligible. The order parameter is then constant for $0 \le \rho \le \rho_0$, with $\Psi 
= \sqrt {N/({\pi} \rho_0^2)}$ and it drops to zero for $\rho > \rho_0$, where $\rho$ is 
the radius in cylindrical coordinates. Minimizing the (interaction) energy, $\rho_0^2 = 
{N \sqrt{e}}/{\pi}$, i.e., $\rho_0$ scales as $\sqrt N$. The corresponding lowest energy 
is $E/N = \mu_{\rm TF} = - 1/(2 \sqrt{e})$, being independent of $N$. Finally, the density 
is also independent of $N$ and equal to $1/\sqrt{e}$.

The Thomas-Fermi profile described above has the problem that it has a discontinuity
at the boundary of the droplet. To cure this, let us introduce the trial order parameter
\begin{eqnarray}
 \Psi(\rho) \propto 
 \sqrt{1 - \tanh \left( \frac {\rho - \rho_0}
{\xi} \right)}.
 \label{trtanh}
\end{eqnarray}
In this state the density is roughly constant for $0 \le \rho \le
\rho_0$, dropping to zero within a length scale of order $\xi$, which is essentially 
the coherence length. In the Thomas-Fermi limit, one expects that $1/(2 \xi^2) \approx 
-\mu_{\rm TF}$, where $\mu_{\rm TF}$ is the chemical potential given above. In this limit 
$\xi$ is of order unity, which is much smaller than the radius of the droplet $\rho_0$,
i.e., $1 \sim \xi \ll \rho_0$. 

In this trial state the kinetic energy per particle scales as $N^{-1/2}$, while the
total energy is given approximately as 
\begin{eqnarray}
\frac E N \approx  - \frac {1} {2 \sqrt{e}} + \frac {0.4425} {\sqrt{N}} + {\cal O} (1/N).
\end{eqnarray} 
Clearly this calculation can be improved, treating $\rho_0$ and $\xi$ variationally. 
The result of this calculation is shown as the dashed line in Fig.\,1. We observe that 
there is excellent agreement between the two curves, essentially for all $N \ge 10$. 
We also stress that, at least for $N \ge 10$ shown in Fig.\,1, the deviation of $\mu$
from $\mu_{\rm TF}$ is very small. Furthermore, on the left plot in Fig.\,2 we compare 
the density profile that is evaluated numerically and variationally, i.e., the one from 
Eq.\,(\ref{trtanh}), for $N=1000$.

\begin{figure}
\includegraphics[width=8.5cm,height=5.5cm,angle=-0]{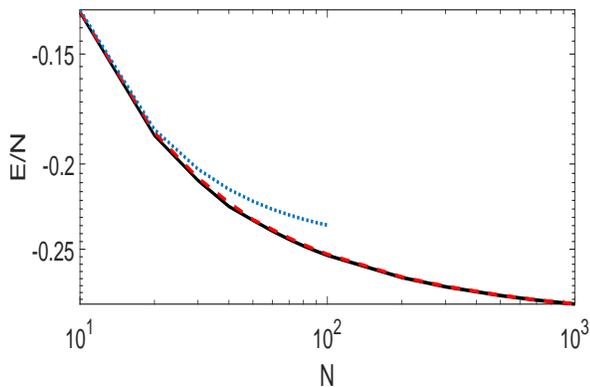}
\caption{The ground-state energy as function of $N$, evaluated from Eq.\,(\ref{nlin}) 
(solid curve), the one that results from the trial order parameter of Eq.\,(\ref{trtanh}) 
(dashed curve) and the one from the Gaussian (dotted curve). Here $\omega = 0$.}
\end{figure}

For small values of $N$ the density of the cloud is Gaussian-like \cite{th3}. Let us thus 
also consider a Gaussian trial function, 
\begin{eqnarray}
\Psi = \frac {\sqrt N} {\sqrt{ \pi \rho_0^2}} e^{-\rho^2/ (2 \rho_0^2)}. 
\end{eqnarray}
Minimizing the energy with respect to $\rho_0$, 
\begin{eqnarray}
\rho_0^2 = \frac {N} {\pi} e^{2 \pi / N},
\end{eqnarray}
while the lowest energy is 
\begin{eqnarray}
  \frac E N =  - \frac 1 4 e^{-2 \pi/N}.
\end{eqnarray}
The dotted curve in Fig.\,1 shows the corresponding energy for various values of $N$. 
As we see in this plot, for small $N$ the agreement is very good, however as $N$ increases, 
it gets worse. 

\section{Droplets under rotation in the absence of any external potential}

\subsection{Vortex excitation}

We now turn to the problem of rotation. Before we examine the yrast state, i.e., the
state of lowest energy for some given angular momentum, we will look for solutions of 
Eq.\,(\ref{nlin}) which have the form $\Psi(\rho, \theta) = f(\rho) e^{i S \theta}$,  
where $S$ is the winding number,
\begin{equation}
 - \frac 1 2 \frac {\partial^2 f} {\partial \rho^2} 
 - \frac {1} {2 \rho} \frac {\partial f} {\partial \rho}
 + \frac {S^2} {2 \rho^2} f + \frac 1 2 \omega^2 \rho^2 f + |f|^2 \ln |f|^2 f = \mu f. 
 \label{nlinnhtr}  
\end{equation}
The solution of the above equation for $|S|=1$ is a vortex state that is located at
the center of the cloud, while for $|S|>1$ we have multiply-quantized vortex states. 
Again, using the method of imaginary-time propagation, we have solved this equation 
numerically. 

In the limit of large $N$, motivated by Eq.\,(\ref{trtanh}) we make the following
ansatz
\begin{eqnarray}
 f(\rho) \propto \frac {\rho} {\sqrt{\rho^2 + \xi_0^2}} \sqrt{1 - \tanh \left( \frac
{\rho - \rho_0} {\xi} \right)}.
 \label{trtanhh}
\end{eqnarray}
Here $\xi$ is again the coherence length that corresponds to the static cloud, and 
$\xi_0$ is a length scale that determines the behaviour of the system close to 
$\rho = 0$.

We stress that for small values of $\rho$, $f$ has to scale linearly with $\rho$, 
in order for the kinetic energy not to diverge. The first factor on the right of 
Eq.\,(\ref{trtanhh}) takes care of this behavior. For $\xi \ll \rho \le \rho_0$, $f$ 
tends to a constant value, as necessary. Thus, the above trial order parameter in a 
sense decouples the ``small" values of $\rho < \xi_0$ from the ``large" values of $\rho 
\approx \rho_0$. Treating $\xi_0$, as well as $\rho_0$ variationally, we minimize 
the energy. The comparison (in the density distribution) between this calculation 
and the full numerical is shown in the right plot of Fig.\,2, for $N=1000$ and $S=1$.

\begin{figure}[t]
\includegraphics[width=9cm,height=5.5cm,angle=-0]{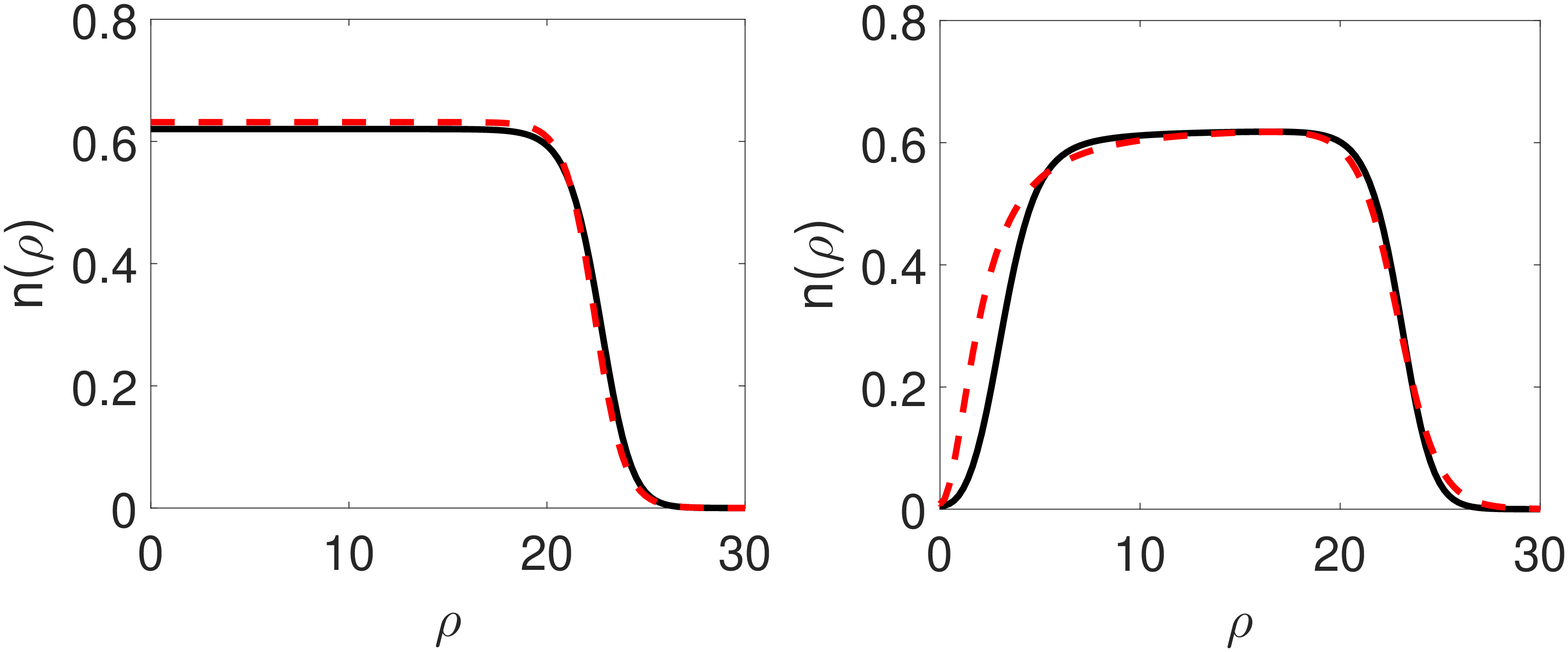}
\caption{The density $n(\rho) = |\Psi(\rho)|^2$ that results from Eq.\,(\ref{nlinnhtr}) 
(solid curve), for $N = 1000$, and the one from the trial order parameter of 
Eq.\,(\ref{trtanh}) (dashed curve), for $S=0$ (left plot), and $S=1$ (right plot). 
Here $\omega = 0$.}
\end{figure}

\subsection{Center of mass excitation}

Up to now, we have restricted ourselves to the case of vortex excitation. If one
wants to give angular momentum to the system, an alternative way to do it is via 
excitation of the center of mass. We stress that since the cloud is self-bound, this 
leaves the interaction energy unaffected and in this respect it is energetically 
favourable. 

Defining some (arbitrary) coordinate system, the cloud of atoms then rotates as 
a whole around the origin. The extra energy is then simply kinetic, which may be 
estimated as $K/N = {N \ell^2}/(2 I)$, where $\ell = L/N$ is the angular momentum 
per atom, and $I$ is the moment of inertia. If the center of mass of the cloud is 
at a distance $r_0$, then $I \approx N r_0^2 + N \rho_0^2$ (neglecting multiplicative 
factors of order unity).

At any value of $r_0$, $K/N \sim 1/N$ for large $N$. Furthermore, since there is no
confining potential, $r_0$ may become infinitely large, i.e., the cloud may escape 
to infinity. In this limit, for some fixed $\ell$, $K/N \to 0$ and therefore the 
yrast energy is in this case the ground-state energy (which is the absolute minimum
of the energy). In other words, in the absence of any trapping potential, the yrast 
state is always the one of center of mass excitation, where the cloud escapes to 
infinity and the energy tends to the ground-state energy.  

We have confirmed the above via minimization of the energy, fixing the angular momentum
$\ell$. To achieve this we use the method of imaginary-time propagation, with the 
following energy functional \cite{kom}, 
\begin{eqnarray}
  \frac {E'} N = \frac E N + \frac C 2 (\langle \hat{\ell} \rangle - \ell_0)^2, 
  \label{lf}
\end{eqnarray}
where $\hat{\ell}$ is the operator of the angular momentum (per atom), $C>0$ is a
constant that has to be sufficiently large (in order for the dispersion relation to 
have a positive curvature) and $\ell_0$ is roughly the value of the angular momentum 
per atom that we wish the droplet to have \cite{rem}. 

\begin{figure}[t]
\includegraphics[width=7cm,height=5cm,angle=-0]{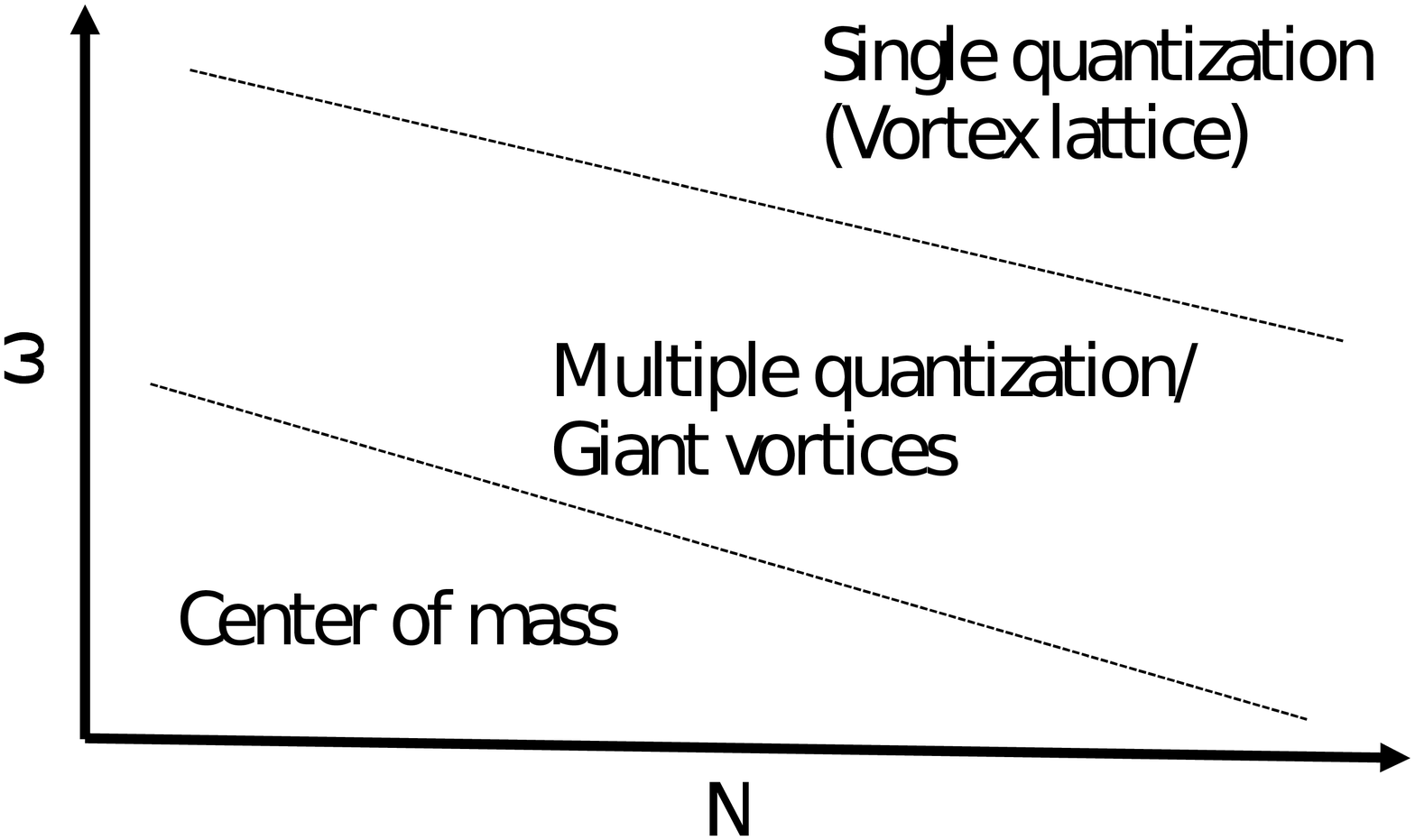}
\vskip2pc
\caption{Schematic phase diagram of a rotating droplet, for some fixed value of $\ell$,
corresponding to Fig.\,4. On the $x$ axis is the number of atoms and on the $y$ axis the 
trap frequency $\omega$. For sufficiently small values of $N$ and $\omega$ we have the 
phase of center of mass excitation. Above it we have the phase of vortices of multiple 
quantization and/or the phase of ghost vortices. The last phase is the top one, where 
we have vortex states of single quantization, which form a vortex lattice, if $\ell$ 
is sufficiently large (since, for $\ell$ being of order unity, there cannot
be any vortex lattice).}
\end{figure}

\begin{figure}
\includegraphics[width=6cm,height=4.5cm,angle=-0]{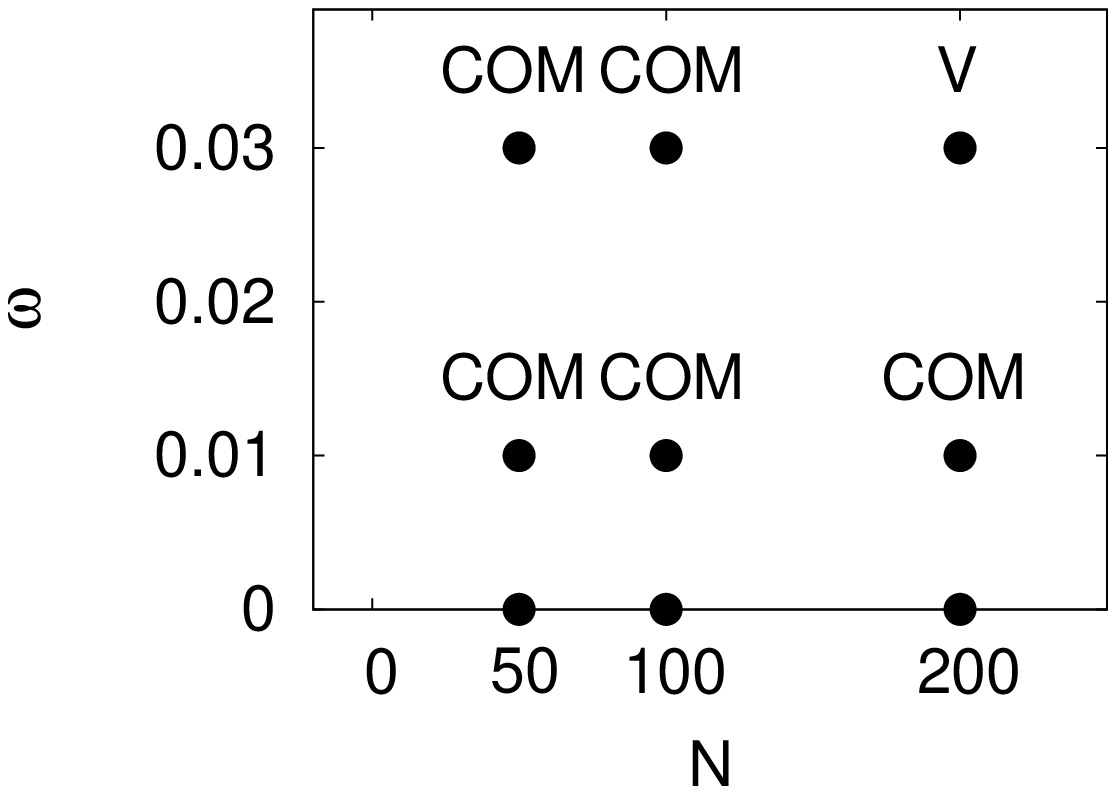}
\includegraphics[width=6cm,height=4.5cm,angle=-0]{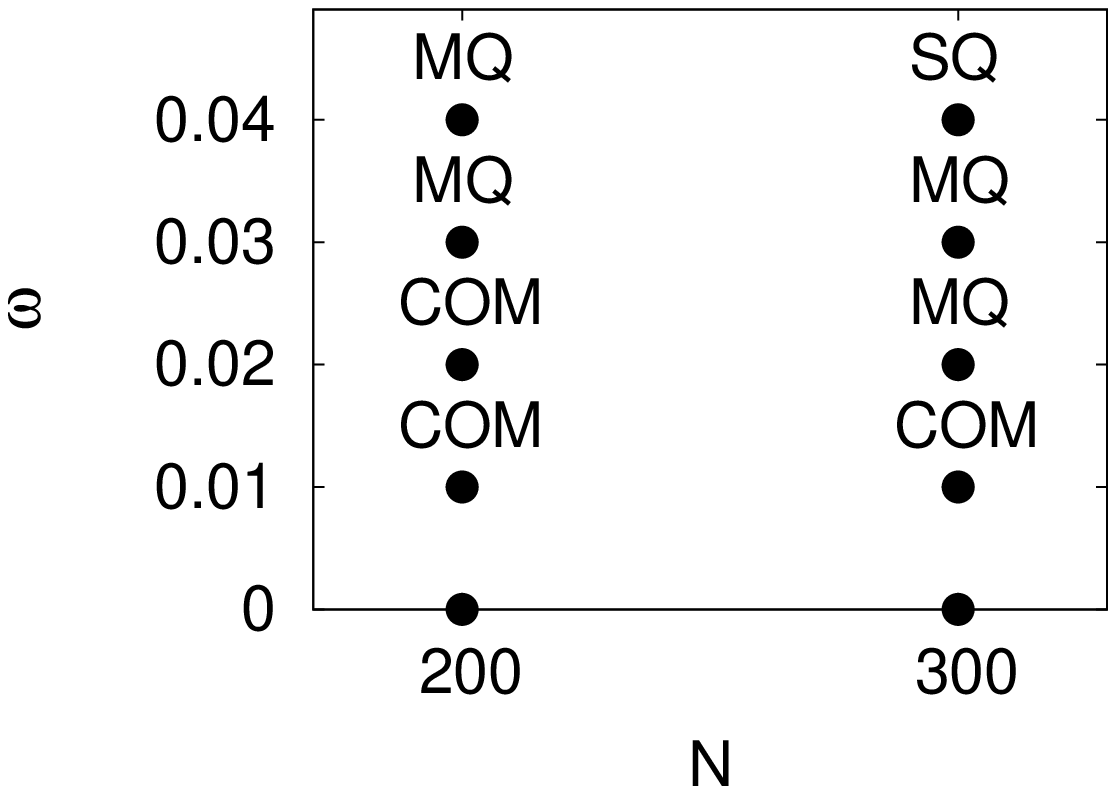}
\includegraphics[width=6cm,height=4.5cm,angle=-0]{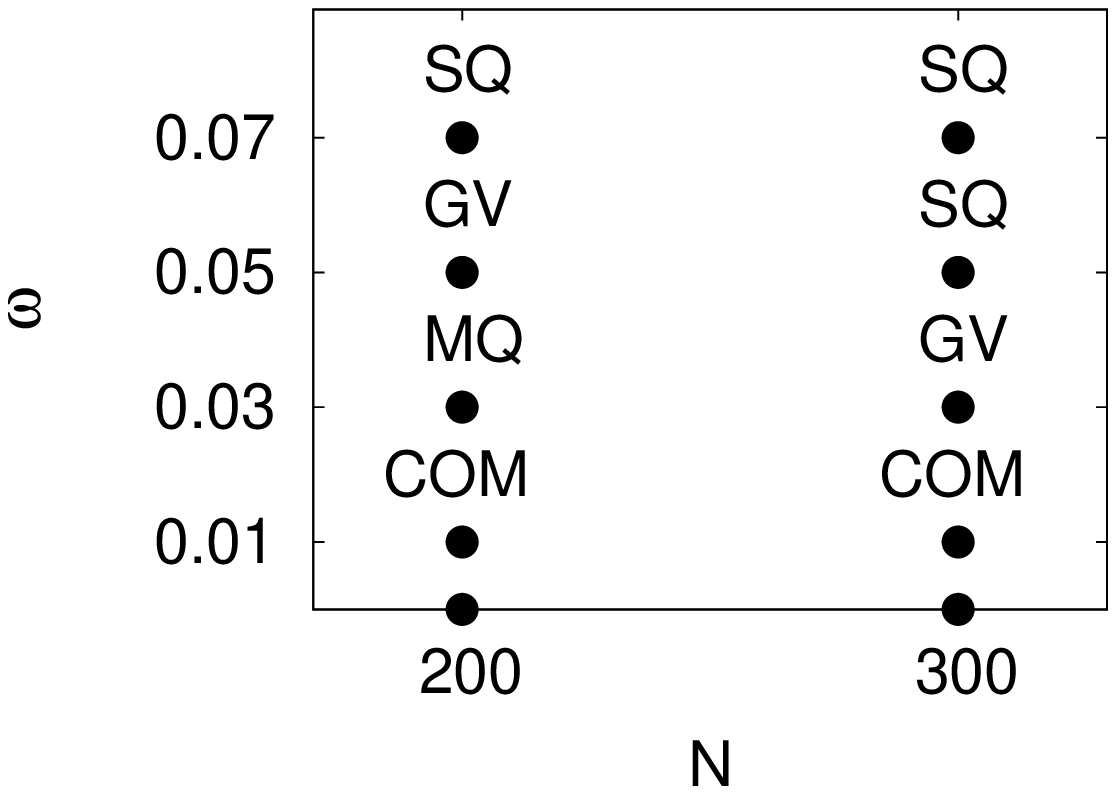}
\vskip2pc
\caption{The phase diagram for a fixed value of $\ell=1$ (top), $\ell=2$ (middle),
and $\ell=3$ (bottom). The ``COM" denotes center of mass excitation, ``V" a single
vortex state, ``MQ" vortices of multiple quantization, ``SQ" singly-quantized vortex 
states, and ``GV" ghost vortices (defined in the text).}
\end{figure}

\section{Droplets under rotation in the presence of a harmonic potential}

According to the arguments presented above, in the absence of any trapping potential
the problem is uninteresting. For this reason, in what follows below we assume that 
there is also a confining potential. To get some insight, it is useful to evaluate the 
energy cost for a vortex state to be located at the center of the cloud and then compare
it with the energy cost for center of mass excitation. 

\subsection{Vortex excitation}

We have evaluated numerically the energy difference $\Delta E_v$ between the state with 
a vortex state at the center of the droplet and the non-rotating state. For $N = 300, 400, 
\dots, 1000$, and for some fixed $\omega$, the energy cost for vortex excitation 
${\Delta E_v}/N$ is fitted by the formula ${\Delta E_v}/N \approx A + B [\ln (N/2)]/N$. 
The logarithmic term is the expected one and it comes from the kinetic energy. This involves 
two terms, the one of which is $\sim \int d^2 \rho/\rho^2$, where the integration extends 
from some lower cut-off, $\xi_0$ in our notation, up to $\rho_0 \sim \sqrt{N}$. Finally, 
$A$ and $B$ are parameters, which depend on $\omega$. 

\subsection{Center of mass excitation}

Turning to the center of mass excitation, in the presence of an external potential the 
situation changes, since $r_0$ is bounded. For any finite value of $\omega$, the cloud 
will spiral outwards, until the restoring force that comes from the confining potential 
will stop it. An estimate for $r_0$ is given by the formula ${l^2}/{r_0^3} \sim \omega^2 
r_0$, or $r_0 \sim \sqrt{\ell} a_0$, where $a_0 = 1/\sqrt{\omega}$ is the oscillator 
length. Therefore, $K/N \approx {\ell^2}/(2 \ell a_0^2 + 2 N \sqrt{e}/\pi)$. The energy 
due to the trapping potential is estimated as $P/N \approx [\omega^2 (r_0^2 + \rho_0^2)]/2 
\approx \omega \ell/2 + \omega^2 N \sqrt{e}/(2 {\pi})$. Fitting the energy difference 
between the rotating state (with center of mass excitation) and the nonrotating for 
$N = 200, \dots, 600$ and a fixed $\omega$, we get that ${\Delta E_{\rm COM}}/N \approx 
C N + D + {\cal O}(1/N)$, in agreement with the estimates given above. Again, $C$ and $D$ 
are parameters, which depend on $\omega$. 

\subsection{Phase diagram}

From the expressions $\Delta E_v/N$ and $\Delta E_{\rm COM}/N$ it is thus clear that in 
the presence of an external potential there is a competition between center of mass and 
vortex excitation. We analyse this effect in more detail below.

\begin{figure}[h]
\includegraphics[width=9.5cm,height=4.5cm,angle=-0]{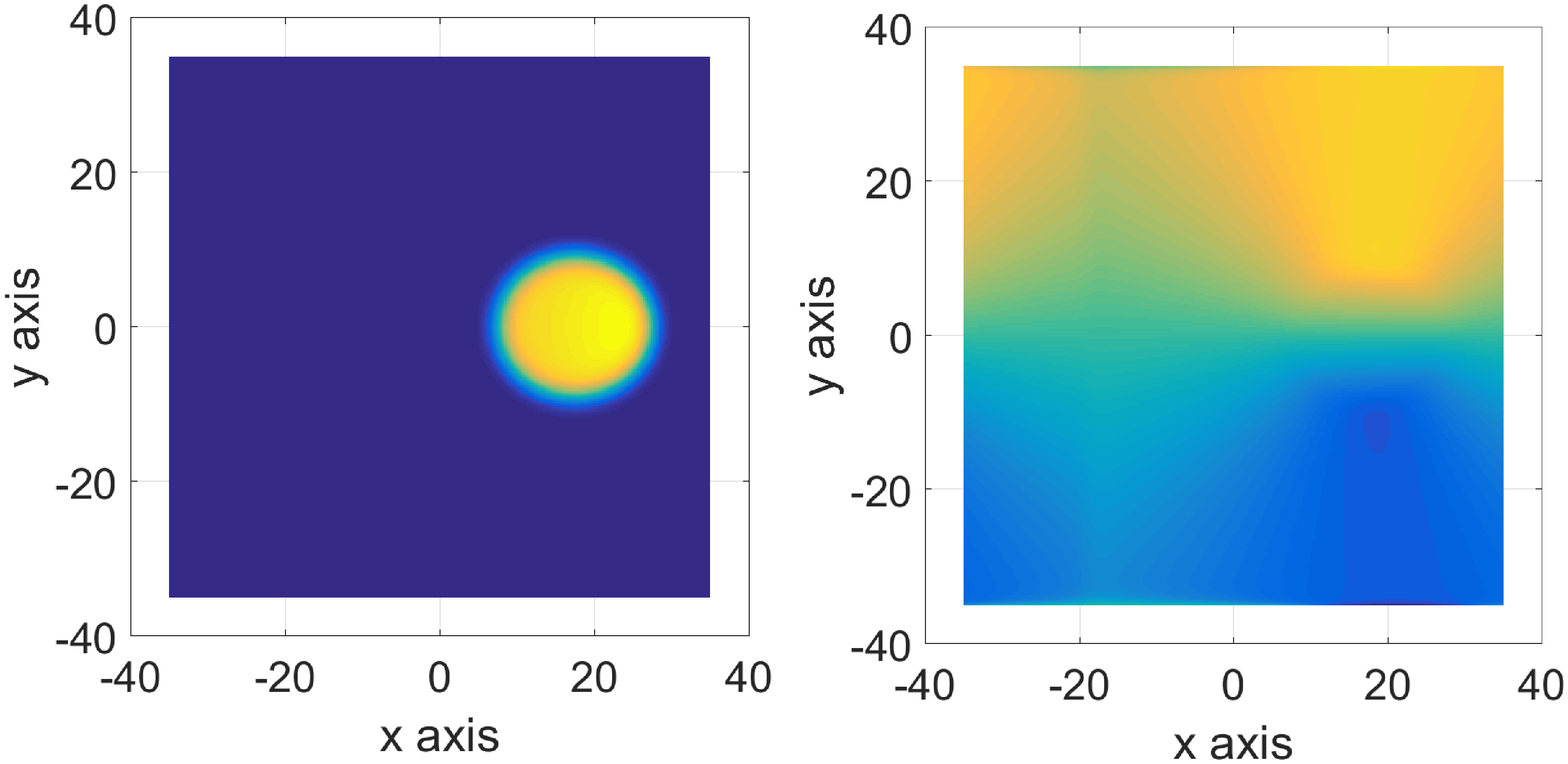}
\includegraphics[width=9.5cm,height=4.5cm,angle=-0]{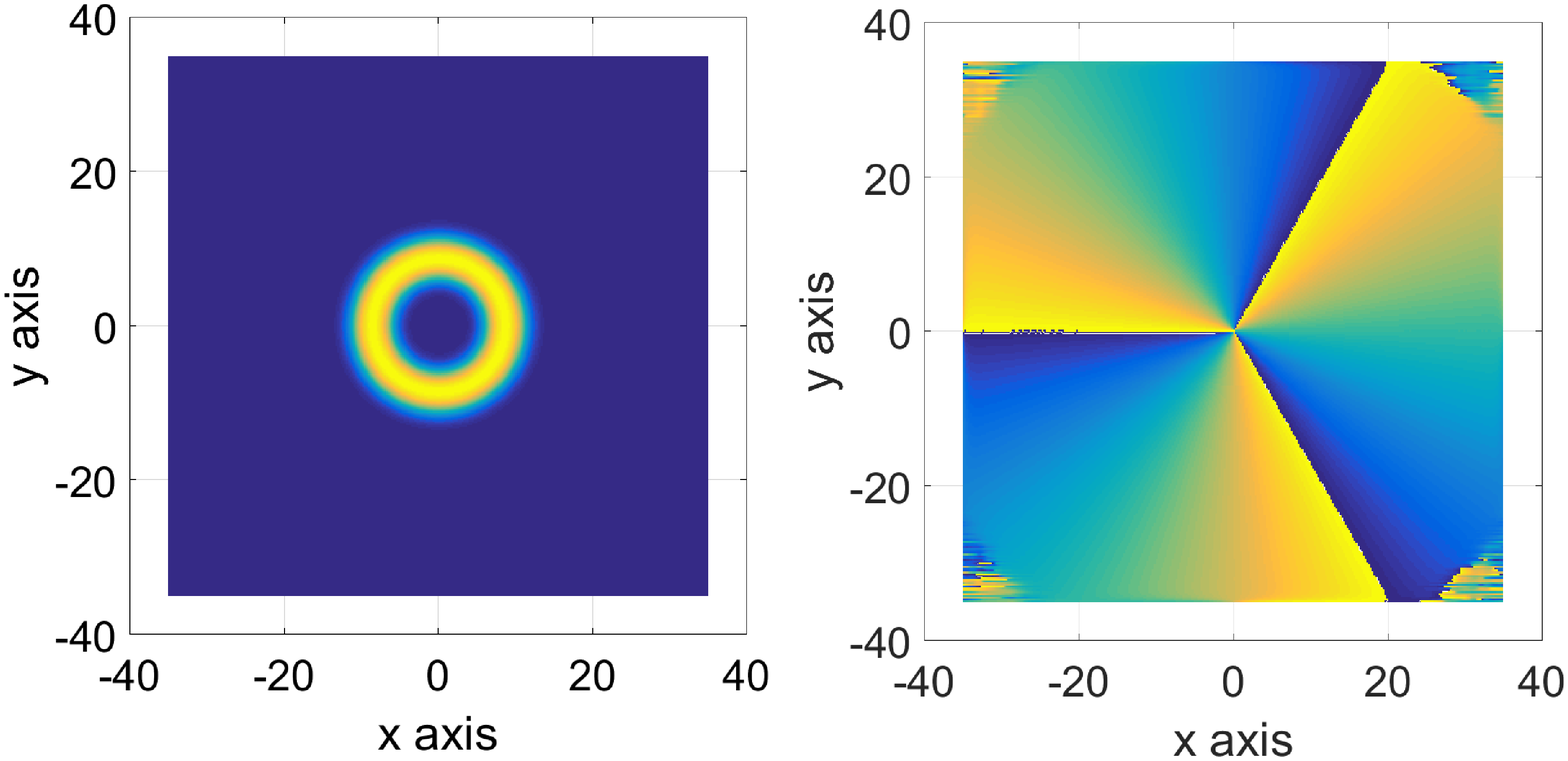}
\includegraphics[width=9.5cm,height=4.5cm,angle=-0]{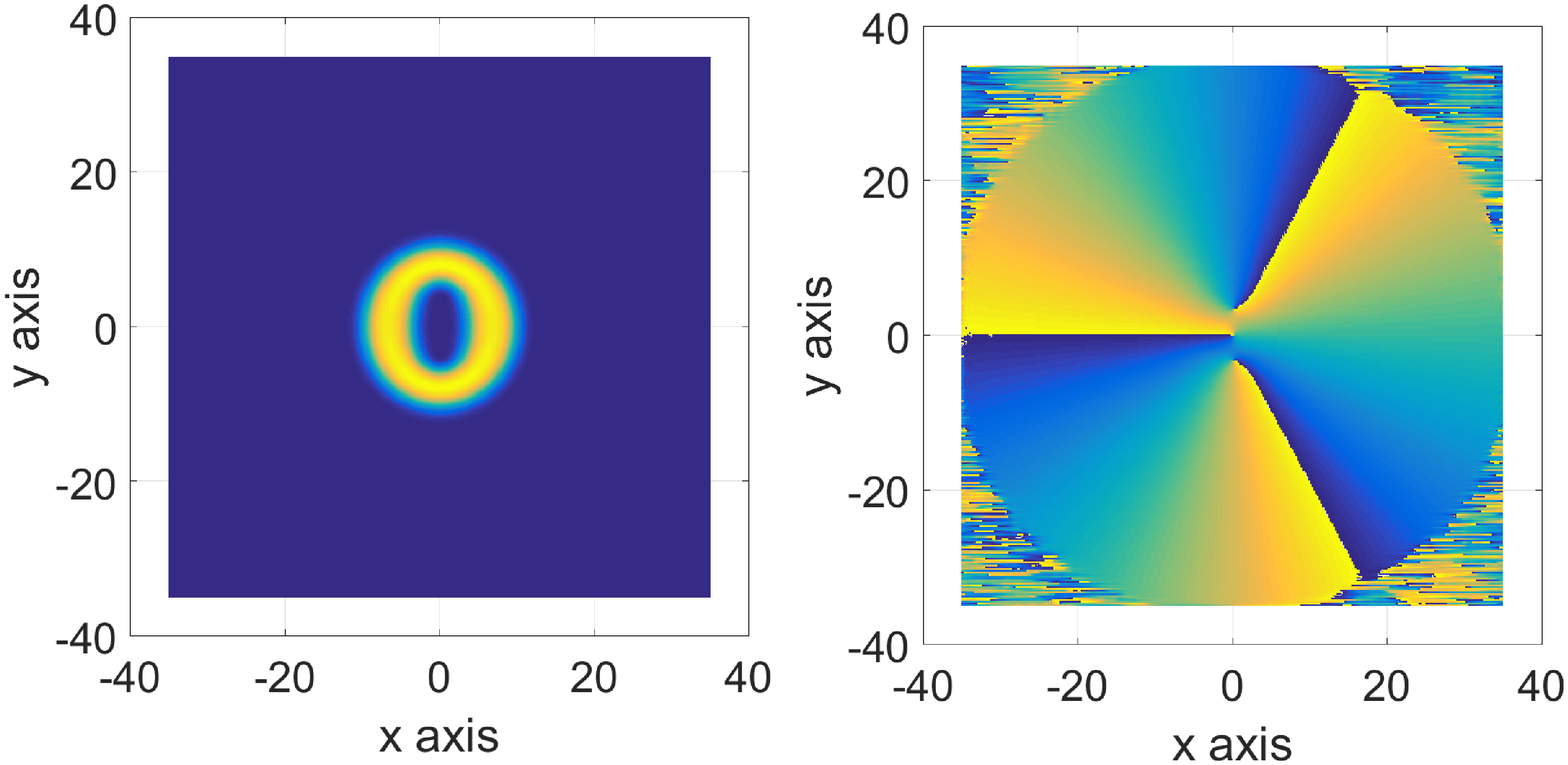}
\includegraphics[width=9.5cm,height=4.5cm,angle=-0]{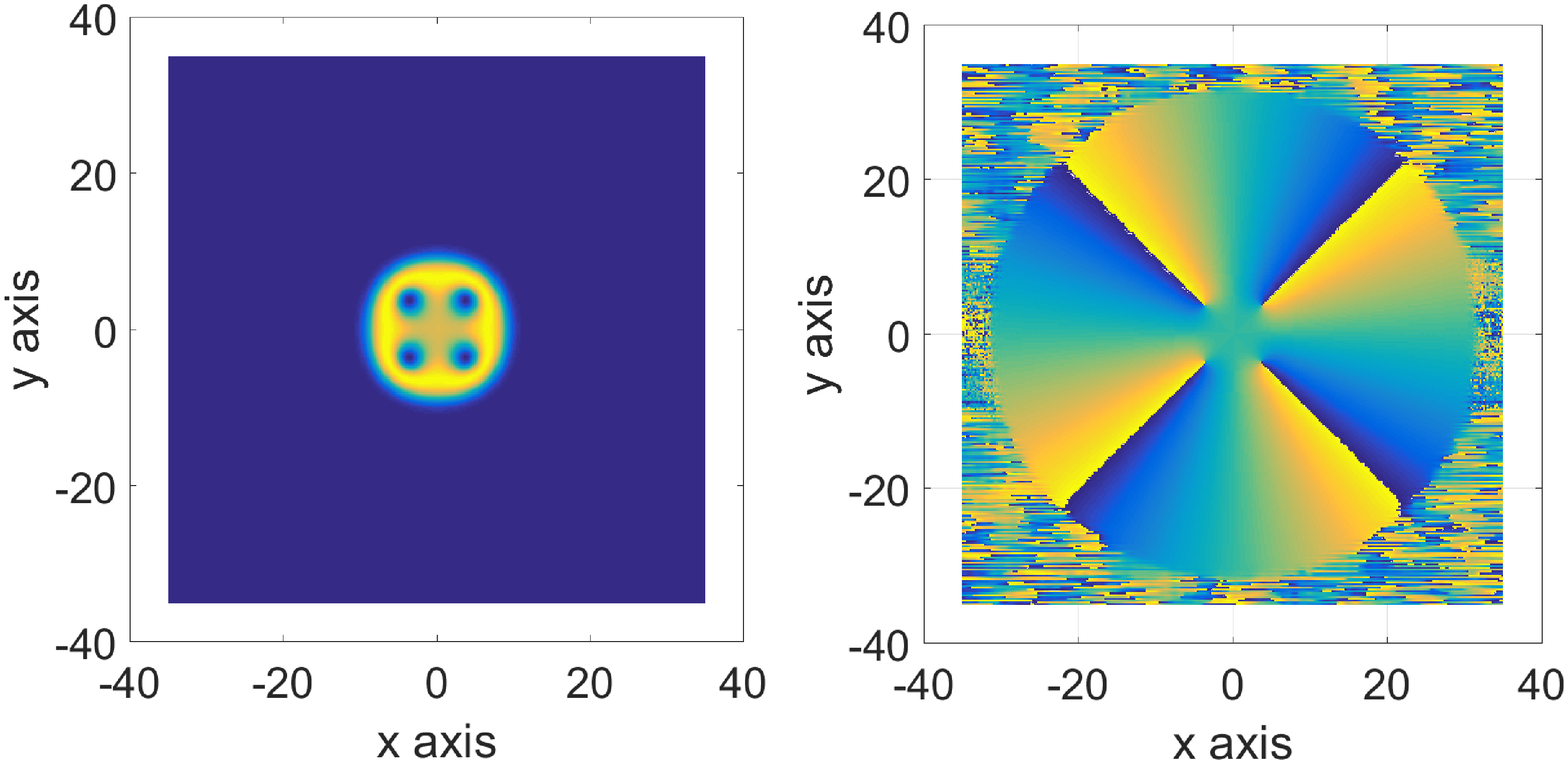}
\caption{The density $n(x,y) = |\Psi(x,y)|^2$ (left column) and the phase of the order 
parameter $\Psi(x,y)$ (right column) from Eq.\,(\ref{lf}), for $N = 200$ and $\ell=3$ in 
the presence of a harmonic potential, with $\omega = 0.01$ (top), 0.03 (second from the 
top), 0.05 (third from the top), and 0.07 (bottom).}
\end{figure}

First of all, we observe that in the absence of the external potential the density is 
sufficiently small and the interaction is purely attractive (since this is the configuration 
of lowest energy). On the other hand, when there is a trapping potential the density may 
become sufficiently high so that the effective interaction may be partly repulsive (where 
the density is high) and partly attractive (where the density is low). This observation is
in the heart of this problem, giving it a rich structure. 

There are three parameters, namely the atom number $N$, the frequency of the trapping 
potential $\omega$, and the angular momentum per atom $\ell$. In principle one may 
investigate the corresponding three-dimensional phase diagram. Instead, we focus below 
on three cases. In all of them we fix $\ell$ (to the value 1, 2 and 3, respectively) 
and we examine the phase diagram, varying $N$ and $\omega$. The results are shown in 
Figs.\,3 and 4.

Figure 3 is a schematic phase diagram, where we show that, quite generally, for 
sufficiently small values of $N$ and $\omega$, the angular momentum is carried via 
center of mass excitation. For higher values of $N$ and/or $\omega$, the system carries 
its angular momentum in the form of vortices of multiple quantization and/or ghost 
vortices (discussed below) \cite{gvss}. For even higher values of $N$ and/or $\omega$ 
we have the ordinary vortex states of single quantization, which form a vortex lattice, 
if $\ell$ is sufficiently large.

In Fig.\,4 we show the results that we have derived by actual numerical simulations, 
for the values of $N$ and $\omega$ shown in each plot. In the upper plot $\ell=1$. 
Here we see the two phases of center of mass excitation and of (single) vortex excitation.
More or less the same picture emerges for $\ell = 2$, which is shown in the middle
of Fig.\,4. Here we also have the possibility of vortices of multiple quantization, 
which compete with the phase of an array of vortices of single quantization. We stress 
that when the interaction is attractive, the exchange interaction favours the vortices 
of multiple quantization \cite{KMP}. On the other hand, both the kinetic energy 
\cite{remark}, as well as the energy due to the trapping potential favour the array of
vortices of single quantization (since these are further away from the origin and 
the density vanishes at the core of each of them), which is energetically favourable.

For $\ell = 3$, shown in the lower plot of Fig.\,4, the situation becomes even more
interesting. Focusing, e.g., on the case $N = 200$ in this plot, as $\omega$ increases 
we have identified four phases. For $\omega = 0.01$ (top plot in Fig.\,5) we have a phase 
of center of mass excitation. For $\omega = 0.03$ (second plot from the top of Fig.\,5) 
we have a triply-quantized vortex state. For $\omega = 0.05$, there are three 
singly-quantized (ghost) vortices, which are located in the region of exponentially small 
density (third from the top of Fig.\,5). Interestingly enough, this region of low density 
is not circular, but rather it is elongated. Finally, for $\omega = 0.07$ we have a regular 
vortex lattice of four singly-quantized vortex states (bottom plot in Fig.\,5). 

Returning to the physical units, let us consider a width of the cloud 
perpendicularly to the plane of motion of the atoms, which is $\sim 0.1$ $\mu$m and 
three-dimensional scattering lengths between the same species $a_{11} = a_{22} = 10$ nm 
and different species $a_{12} = - 10.1$ nm. Then, the two-dimensional density of the 
droplets on their plane of motion (in the absence of any trapping potential) is $\sim 
5 \times 10^8$ cm$^{-2}$, which is much smaller than $a^{-2}$, as it should. Furthermore, 
the corresponding three-dimensional density is $\sim 5 \times 10^{13}$ cm$^{-3}$. Finally, 
for a number of atoms $10^3$, the typical size of the droplet is $\sim 10$ $\mu m$, which 
corresponds to an $\omega \sim 10$ Hz. The actual value of $\omega$ should be of this 
order, or smaller, especially if one wants to realize the phase of center of mass 
excitation.

\section{Summary and conclusions}

To conclude, while in the past it has become possible to tune the sign of the
effective interaction using the method of Feshbach resonances, in the present system 
this becomes possible by tuning either the atom number, or the strength of the 
confining potential. It is even more interesting that the sign of the nonlinear term 
may vary spatially, depending on the value of the density. 

More specifically, for low atom numbers and/or weak confinement, the physics of the
system is determined by the attractive sign of the nonlinear term, and in the case 
of rotation the yrast state involves center of mass excitation. In the opposite limit 
of high atom numbers and/or strong confinement, the nonlinear term becomes repulsive 
and the yrast state involves vortex excitation. 

An interesting scenario which results from this study, is the possibility of center-of-mass 
motion for some range(s) of $\ell$, and vortex excitation for other range(s) of $\ell$. 
This would give a positive curvature in the dispersion relation in the first case and a 
negative curvature in the second, resembling, in a sense the phonon-roton spectrum. Such 
a dispersion relation would also affect the ability of the system to support persistent 
currents.

The system we have investigated is ideal for the study of superfluid states. It is
remarkable that by simply tuning either the atom number, or the strength of the 
confining potential, one is able to switch between the phases we have found. Similar 
phases have been found (separately) in other superfluid Bose-Einstein condensed gases, 
as in a harmonic potential with an effective attractive, or repulsive interaction 
\cite{sara}, and in a quadratic-plus-quartic potential \cite{kjb}. 

For the parameters we have considered, we have identified four phases, however the 
existence of more phases is not excluded. We stress that the phase diagrams we have 
found are generic. On the other hand, it is clear that more theoretical and experimental 
work is necessary, in order to get a more complete picture of the possible phases, and 
of the order of the transition between them.

\acknowledgements The authors wish to thank Stephanie Reimann, Elife Karabulut, and
Manolis Magiropoulos for useful discussions.


\begin{thebibliography}{55}
\bibitem{rem} Within the yrast problem that we consider here, we are interested in the 
global minimum of the energy for some given value of the angular momentum. While converged
solutions of the method of imaginary-time propagation definitely correspond to local minima
of the energy, one can never be certain that the global minimum has been found. We should
also add that such local minima are not uninteresting, since they are relevant also 
experimentally.
\end{thebibliography}
\end{document}